# Geometric Stability of Pseudo-plane Ideal Flows

Che Sun[1,2]

[1] Institute of Oceanology, Chinese Academy of Sciences, Qingdao, China

[2] National Laboratory for Marine Science and Technology, Qingdao, China

(Corresponding email: csun@qdio.ac.cn)

**Abstract.** Geometric analysis of steady pseudo-plane ideal flow reveals a fundamental relation between vertical coherence and streamline topology. It shows vertical alignment only exists in straightline jet and circular vortex. A geometric stability theory is then developed as a structural stability in contrast to traditional dynamical stability, suggesting all vertical-aligned pseudo-plane flows are geometrically stable and all non-aligned non-straightline flows are unstable. It provides a topological explanation for the tendency of axisymmetrization and vertical alignment observed in vortex experiments.

# 1. Introduction

Pseudo-plane ideal flow is a quasi-three-dimensional type of Euler flow with vertically varying horizontal velocities and no vertical velocity. Vortex stretching, a main source of vorticity fluctuation in turbulence, is absent in pseudo-plane flow. This makes pseudo-plane velocity a basic laminar topology for stably stratified fluid. Indeed, laboratory experiments have shown that freely decaying turbulence in stratified fluid tends to collapse into pancake layers where vertical velocity is negligible (*Riley and Lelong* 2000). In atmosphere and ocean, effect of pseudo-plane flow is manifested by the dominance of Gravest Empirical Mode (*Sun and Watts* 2001, *Sun* 2005), which motivates *Sun* (2008, hereafter Sun08) to develop a steady pseudo-plane flow (PIF) model to study baroclinic coherent structure. A Lie-symmetry analysis of the model was performed by *Lou et al.* (2012).

The exact solutions to the PIF model have been used to validate the geometric properties of pseudo-plane flows (*Sun* 2016). The stability of these steady solutions, however, has not been investigated. As explained by *Landau and Lifshitz* (1959), not every exact solution can occur in reality, and those which do occur must not only obey the fluid dynamics equations but also be dynamically stable. Hydrodynamic stability therefore becomes an essential problem in fluid mechanics. It examines the temporal evolution of a small perturbation on a steady laminar flow, and the basic flow is said to be unstable if the perturbation grows in amplitude with time (*Lagnado et al.* 1984, *Craik and Criminale* 1986, *Dritschel* 1990, *Salhi et al.* 1996, *Shapiro and Fedorovich* 2012).

Such conventional stability analysis is conducted in time domain and often limited to two-dimensional basic flow. It can not be applied to a steady-state model like PIF. For steady pseudo-plane flows, we propose a geometric stability concept in spatial domain based on exact solutions of the nonlinear PDE system. The approach would extend stability analysis to three-dimensional flows and yield a topological theory for the phenomenon of vortex axisymmetrization and vertical alignment (*Polvani* 1991, *Nof and Dewar* 1994, *Viera* 1995, *Sutyrin et al.* 1998, *McWilliams et al.* 1999, *Reasor and Montgomery* 2001).

For the purpose of wider application, Section 2 reviews the baroclinic theory of Sun08 and extends it to general pseudo-plane ideal flows. A geometric stability theory is then developed in Section 3 and applied to a series of quadratic flows in Sections 4. Relation between geometric stability and hydrodynamic stability is discussed in Section 5.

## 2. Geometric properties for pseudo-plane flows

The PIF model of Sun08 represents a laminar end state of stratified turbulence and was derived from the classical stratified-turbulence model by taking the steady-state limit of vanishing vertical velocity under Boussinesq approximation. Latter *Sun* (2016) gave an alternate derivation directly from the Euler equations in a rotating frame with steady angular velocity $\Omega$. The PIF model consists of five PDE

$$uu_x + vu_y - fv = -p_x \tag{1}$$

$$uv_x + vv_y + fu = -p_y \tag{2}$$

$$\rho = -p_z \tag{3}$$

$$u_x + v_y = 0 \tag{4}$$

$$u\rho_x + v\rho_y = 0 \tag{5}$$

where $u(x,y,z)$ and $v(x,y,z)$ are horizontal velocities, $f = 2\Omega$ is the Coriolis parameter, $p$ is pressure perturbation divided by a mean density $\rho_0$, and $\rho$ is density perturbation scaled by $\rho_0/g$. Background pressure $\bar{p}(z)$ and density $\bar{\rho}(z)$ in hydrostatic balance are neglected because they exert no dynamical effect on pseudo-plane flows. Pressure and density perturbations superposed upon the background field must have horizontal variations.

A streamfunction $\psi$ can be introduced as $u = -\psi_y$, $v = \psi_x$ according to Eq.(4). Four Lagrangian invariants of the PIF model have been identified, including density, mechanical energy, potential vorticity and vertical vorticity. The Boussinesq approximation ignores density variations in the horizontal momentum equations and the continuity equation (*Spiegel and Veronis* 1960). It leads to an absence of solenoidal term in the vertical vorticity equation and renders vertical vorticity $\zeta = \nabla^2 \psi$ a Lagrangian invariant in pseudo-plane flows, i.e., $J(\psi, \zeta) = 0$. Meanwhile mechanical energy is also conserved along streamline, i.e., $J(\psi, p + K) = 0$, where kinetic energy $K = (u^2 + v^2)/2$ can represent speed.

Several geometric concepts for pseudo-plane motions are defined in the following.

i) **Pseudo-plane flow:** a flow with zero vertical velocity and vertically-varying horizontal velocity, i.e., $\mathbf{u} = [u(x,y,z), v(x,y,z), 0]$. If horizontal velocity is vertically invariant, i.e., $u_z = v_z = 0$, the flow is **plane flow**. For a plane flow, vertical differentiation of Eqs. (1-2) gives $\rho_x = \rho_y = 0$, which means no density perturbation and isopycnals are flat. Pressure perturbation may exist in plane flow, but shall be vertically invariant.

ii) **Baroclinic flow**: a pseudo-plane flow is baroclinic if isobaric surfaces and isopycnal surfaces do not coincide ($\nabla \rho \times \nabla p \neq 0$). Isobaric surfaces in a baroclinic flow can not be flat, otherwise $p_x = p_y = 0$ and vertical differentiation would give $\rho_x = \rho_y = 0$; A flow with flat isopycnals is not baroclinic either, because it does not store the baroclinic part of available potential energy.

iii) **Equivalent-barotropic flow**: a pseudo-plane flow is EB if its streamlines on each plane align vertically or, equivalently, if the horizontal velocity vector does not change direction vertically (Figure 1 is a non-EB example). Plane flow is an extreme case of EB formulation.

Unlike the EB concept in meteorology, the definition here is purely a geometric characterization of vertical alignment and is not related to geostrophic dynamics.

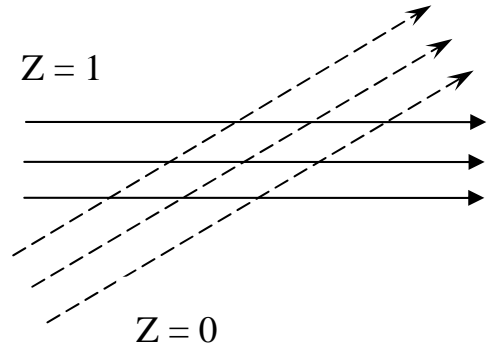

**Figure 1**. Non-EB straightline jet with direction varying vertically.

Because baroclinic flows always have horizontally varying pressure, the six geometric properties in Sun08 can be rephrased to apply to general pseudo-plane flows. Unless being specified, the properties are applicable to both homogeneous fluid ($\rho = 0$) and stratified fluid, and both rotating frame and non-rotating frame ($f = 0$) as well.

**Lemma 1**. *A streamfunction field $\psi(x, y, z)$ is EB if and only if $J(\psi, \psi_z) = 0$ holds everywhere.*

**Lemma 2**. *For steady pseudo-plane or plane flows, if pressure or speed is constant along streamlines, then streamlines on a horizontal plane must be parallel straight lines or concentric circles.*

**Lemma 3**. *A pseudo-plane flow with streamline slope $\eta$ ($v = \eta u$) satisfying $\eta \eta_x - \eta_y = 0$ everywhere must assume parallel straight lines or concentric circles at each plane, and vice versa.*

**Theorem 1**. *If a steady pseudo-plane flow is EB, it belongs to constant-speed flow and appears as straightline jet or circular vortex.*

**Theorem 2**. *A steady pseudo-plane flow with straight-line or circular streamlines must be EB if pressure perturbation exists.*

**Theorem 3**. *A steady pseudo-plane flow is EB if and only if its helicity density vanishes.*

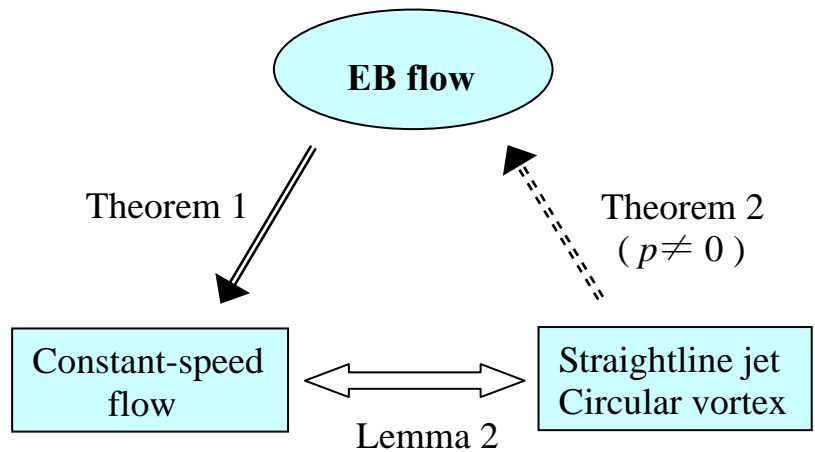

**Figure 2.** Schematic illustration of interrelationships among key geometric properties. The dashed arrow denotes a conditional relation.

These geometric properties, as illustrated in Figure 2, reveal a fundamental relation between vertical coherence (EB flow) and horizontal topology (streamline pattern) in ideal fluid, which is summarized as follows: for steady pseudo-plane flows, only straightline jet and circular vortex can have vertical alignment (EB flow) formulation, and all other flow patterns are vertically unaligned and devoid of vertical coherence. The pressure condition of Theorem 2 implies that straightline jet is always EB in rotating frame and circular vortex is always EB in non-rotating frame, which will be verified by exact solutions in Section 4.

Such low-dimensional geometric properties, elegant as they are, do not tell us the existence and stability of pseudo-plane exact solutions, particularly those non-EB ones. To examine the stability of exact solutions is the subject of the present study.

## 3. Geometric stability theory

We now study the stability of pseudo-plane flows based on the so-called Gilbarg's problem (*Gilbarg* 1947, *Munk and Prim* 1947), which is formulated as follows: under what conditions is the flow dynamics (velocity field) uniquely determined by its geometry (streamline pattern)? If two steady flows have the same streamline pattern, their velocity vectors at each point have the same direction, i.e.,

$$\hat{\mathbf{u}} = F(x, y, z)\mathbf{u}$$

Flow **u** is unique if all other flows with the same streamline pattern have velocities in proportion to it, that is, if $F$ is constant. For homogeneous ideal fluid in non-rotating frame, *Gilbarg* (1947) proved that all steady plane flows are unique except for those with constant speed on streamlines (in rotating frame, however, we will see that not all constant-speed flows are non-unique). We set to extend Gilbarg's problem to pseudo-plane flows and propose a definition of geometric stability as follows.

**Definition** (Geometric stability). Flow **u** is non-unique and geometrically stable if there exists $\hat{\mathbf{u}}$ with horizontally varying $F$. Otherwise the flow is unique and geometrically unstable. For unstable cases, flow **u** is strictly unique if $F$ is constant for all $\hat{\mathbf{u}}$, and horizontally unique if there exists $\hat{\mathbf{u}}$ with $F = F(z)$.

The concept of geometric stability belongs to structural stability and is thus equivalent to the non-uniqueness of a pseudo-plane flow. It is based on the following consideration: 1) in a freely evolving flow system, the presence of viscous damping causes the kinetic energy to decrease with time; 2) the flow velocity can not decrease in a globally proportional way, because viscous damping arises from boundary layer and is not spatially uniform; 3) only a non-unique flow can accommodate such non-proportional velocity changes without altering its flow pattern; 4) the velocity field of a unique flow, if decreases, must be in global proportion. We consider a unique flow to be geometrically unstable and less likely to maintain its geometric pattern during a viscous decay.

Assuming ($\hat{u}, \hat{v}, \hat{p}, \hat{\rho}$) is a new solution to the PIF model with the same streamlines as ($u, v, p, \rho$), we substitute

$$\hat{u} = F(x,y,z)u, \quad \hat{v} = F(x,y,z)v \qquad (6)$$

into Eq.(4) and obtain

$$0 = \hat{u}_x + \hat{v}_y = (uF)_x + (vF)_y = uF_x + vF_y$$

which leads to

$$J(\psi, F) = 0 \tag{7}$$

Meanwhile substituting (6) into Eqs(1-2) gives

$$-\hat{p}_x = -F^2 p_x + f v(F^2 - F)$$
$$-\hat{p}_y = -F^2 p_y - f u(F^2 - F) \tag{8}$$

Eliminating $\hat{p}$ in Eqs.(8) by cross-differentiation and using Eq.(4) and Eq.(7), we obtain

$$J(F, p) = 0 \tag{9}$$

If $F$ is horizontally non-uniform, combining (7) and (9) yields

$$J(\psi, p) = 0 \tag{10}$$

Because mechanical energy is a Lagrangian invariant in the PIF model, equality (10) means speed is constant on streamline. It suggests that constant-speed flow (CSF) can be non-unique, but $F$ for non-constant-speed flow (NCSF) must be horizontally uniform. Further constraint comes from the density conservation condition. Substituting (6) and (8) into Eq.(5) gives

$$0 = \hat{u}\,\hat{p}_{xz} + \hat{v}\,\hat{p}_{yz} = F^2 [2F_z J(\psi, p) - f(F-1) J(\psi, \psi_z)]$$

Therefore

$$2F_z J(\psi, p) = f(F-1) J(\psi, \psi_z) \tag{11}$$

CSF in non-rotating frame is non-unique as it automatically satisfies both (10) and (11). In rotating frame, however, equality (11) requires CSF to be EB, i.e., $J(\psi, \psi_z) = 0$. Non-EB CSF in rotating frame is strictly unique because equality (11) requires $F=1$.

NCSF satisfies $J(\psi, p) \neq 0$ and must have horizontally uniform $F$, i.e., $F = F(z)$, otherwise it contradicts with condition (10). In non-rotating frame, equality (11) requires the NCSF to have $F_z = 0$, which means $F$ is constant and the flow is strictly unique. In rotating frame, we define a $G$ function for NCSF as

$$G(x, y, z) = \frac{f}{2} \frac{J(\psi, \psi_z)}{J(\psi, p)} \qquad (12)$$

and equality (11) becomes

$$F_z = (F - 1) G \qquad (13)$$

If $G$ is horizontally uniform, NCSF is horizontally unique and integration of (13) gives

$$F(z) = c \exp(\int G(z) dz) + 1 \qquad (14)$$

If $G$ varies horizontally, equality (13) can not hold unless $F = 1$, which means the NCSF is strictly unique.

Table 1. Classification of pseudo-plane flows and their geometric stability.

|  |  | Non-EB | EB |
|---|---|---|---|
| Non-rotating Fluid | CSF | ( I ) stable | ( II ) stable |
|  | NCSF | ( III ) unstable |  |
| Rotating fluid | CSF | ( IV ) unstable | ( V ) stable |
|  | NCSF $G(x, y, z)$ | ( VI ) unstable |  |
|  | NCSF $G(z)$ | (VII) horizontally unstable |  |

We thus complete the analytical derivation of geometric stability for pseudo-plane flows. As summarized in Table 1, CSF is the necessary condition for geometrically stable flows and NCSF is the sufficient condition for geometrically unstable flows. Table 1 also reflects the requirement of Theorem 1 that EB flows must be CSF and all NCSF must be non-EB. It means that EB flows always satisfy the stability conditions (10)-(11). Therefore EB is the sufficient condition for geometrically stable flows and non-EB is the necessary condition for geometrically unstable flows. The non-EB type of CSF in rotating frame is geometrically unstable, in contrast with the Gilbarg's problem in non-rotating frame where CSF is always stable.

This connection between vertical coherence (EB or non-EB) and geometric stability enables us to extend Theorem 4 of Sun08 to general pseudo-plane flows:

**Theorem 4**. *All EB flows are geometrically stable. Non-EB flows, except for straightline jet in non-rotating frame, are strictly or horizontally unstable.*

It shows non-EB is a hallmark feature of unstable pseudo-plane flows. Because Theorem 1 requires that EB flow is either straight-line jet or circular vortex, Theorem 4 leads to the following theorem.

**Theorem 5**. *Straightline jet and circular vortex in non-rotating frame, or their EB forms in rotating frame, are geometrically stable. All other plane* *or pseudo-plane flows are geometrically unstable.*

The theorem implies that a pseudo-plane flow with closed streamlines, if being geometrically stable, must assume a vertically aligned formulation. It explains the vertical coherence of vortices observed by *McWilliams et al.* (1999), if the laminar end state of a decaying turbulence is assumed to be in stable pseudo-plane motions.

## 4. Application to quadratic flows

Quadratic flow is a basic type of pseudo-plane flows in which velocity components vary linearly with spatial coordinates. The velocity gradient and strain rate are spatially uniform. Due to this unique property, quadratic flow has been commonly used as mean flow in hydrodynamic stability analysis (see the references in Section 1).

The complete set of quadratic solutions to the PIF model is listed below, of which S1-S6 are in non-rotating frame and S6-S13 are in rotating frame. Their geometric stability types, according to Table 1, are denoted in parenthesis. The derivation of these exact solutions is given in *Sun* (2017) and can be easily verified by Maple software.

**(S1)** Straightline jet (Type I, Type II)

$$\psi = [a(z)x + b(z)y + h(z)]^2$$
$$u = -2b(ax+by+h), \quad v = 2a(ax+by+h)$$
$$p = 0, \quad \rho = 0$$

The straightline jet is generally non-EB as its direction changes with depth (Figure 1). The non-EB formulation does not violate Theorem 2 because pressure perturbation is absent. The jet becomes EB when $b/a$ is constant or one of $a$ and $b$ is zero. Being EB or not, straightline jet as a non-rotating CSF is always geometrically stable.

**(S2)** Circular vortex (Type II)

$$\psi = a(z)(x^2 + y^2)$$
$$u = -2ay, \quad v = 2ax$$
$$p = 2a^2(x^2 + y^2), \quad \rho = -4aa'(x^2 + y^2)$$

The solution is a baroclinic circular vortex. In non-rotating frame, centrifugal acceleration can only be balanced by horizontal pressure gradient. The presence of pressure perturbation requires this circular vortex to be EB according to Theorem 2, and it is geometrically stable (Figure 3a).

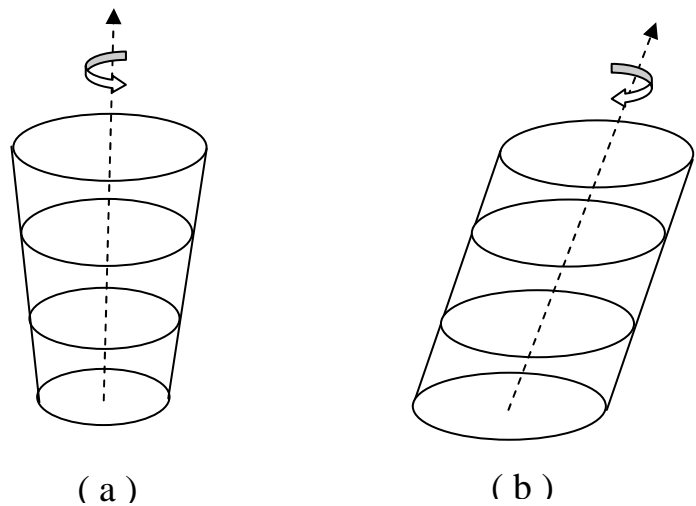

(a)　　　　(b)

**Figure 3**. (a) EB circular vortex. (b) non-EB inertial circular vortex.

**(S3) Elliptical or hyperbolic flow in pulsation mode (Type III)**

$$\psi = a(z)x^2 + \frac{c}{a(z)}y^2$$

$$u = -\frac{2c}{a}y, \quad v = 2a\,x$$

$$p = 2c(x^2 + y^2), \quad \rho = 0$$

The non-EB flow is elliptical for $c > 0$ and hyperbolic for $c < 0$.

**(S4) Parabolic flow (Type III)**

$$\psi = a(z)[x + x_0(z)]^2 + \frac{c\,y}{a(z)}$$

$$u = -c/a, \quad v = 2a(x + x_0)$$

$$p = 2cy, \quad \rho = 0$$

The flow is non-EB as its symmetry line $x = -x_0$ shifts horizontally.

**(S5) Hyperbolic flow in rotary mode (Type III)**

The shape-preserving hyperbolic solution

$$\psi = (c_1 \cos^2 Z - c_2 \sin^2 Z)x^2 + (c_1 \sin^2 Z - c_2 \cos^2 Z)y^2$$
$$\pm (c_1 + c_2)xy \sin 2Z$$

$$p = -2c_1 c_2 (x^2 + y^2), \quad \rho = 0$$
$$Z = Z(z), \quad c_1 > 0, \quad c_2 > 0$$

is geometrically unstable. If $c_1 = c_2$, the solution takes the simple form $\psi = a r^2 \cos(2Z \pm 2\theta)$ in cylindrical coordinates, which is a pseudo-plane version of the dipolar strain field frequently used in vortex studies. Such strain flows are inherently unstable.

**(S6) Elliptical vortex in rotary mode (Type III)**

$$\psi = (c_1 \cos^2 Z + c_2 \sin^2 Z)x^2 + (c_1 \sin^2 Z + c_2 \cos^2 Z)y^2$$
$$\pm (c_1 - c_2)xy \sin 2Z$$

$$p = 2c_1 c_2 (x^2 + y^2), \quad \rho = 0$$

The flow is non-EB and NCSF, and belongs to the same solution type as S3, S4 and S5. They are all geometrically unstable.

**(S7) Inertial circular vortex with skew center (Type IV)**

$$\psi = -\frac{f}{2}[x + x_0(z)]^2 - \frac{f}{2}[y + y_0(z)]^2$$
$$p = 0, \quad \rho = 0$$

The non-EB solution describes free inertial motion in the absence of horizontal pressure gradients (Figure 3b). It does not violate Theorem 2 because the centrifugal force here is balanced by the Coriolis force, leaving the pressure field unperturbed. Equality (11) for this non-EB CSF solution requires $F=1$, which means it is strictly unique and geometrically unstable.

**(S8) Baroclinic circular vortex (Type V)**

$$\psi = a(z)(x^2 + y^2)$$
$$u = -2ay, \quad v = 2ax$$
$$p = a(2a + f)(x^2 + y^2)$$
$$\rho = -a'(4a + f)(x^2 + y^2)$$

This EB circular vortex is CSF and geometrically stable (Figure 3a).

**(S9) Baroclinic straightline jet (Type V)**

$$\psi = [a(z)y + b(z)]^2$$
$$u = -2a(ay + b), \quad v = 0$$
$$p = f\psi, \quad \rho = -f\psi_z$$

In rotating fluid a straightline jet always incurs horizontal pressure variation and must be EB per Theorem 2. The EB flow is geometrically stable.

**(S10) Inertial elliptic or hyperbolic flow (Type VI)**

$$\psi = -\frac{f}{2}x^2 + c[y+y_0(z)]^2$$
$$u = -2c(y+y_0), \quad v = -fx$$
$$p = -\frac{f}{2}(2c+f)x^2, \quad \rho = 0, \quad c \neq -\frac{f}{2}$$

It describes an elliptic ($c<0$) or hyperbolic ($c>0$) flow with skew center. This non-EB NCSF flow is strictly unique because its *G*-function varies with *y*:

$$G(y,z) = -\frac{f y_0'}{2(y+y_0)(2c+f)}$$

**(S11) Inertial parabolic flow (Type VII)**

$$\psi = -\frac{f}{2}x^2 + a(z)y$$
$$u = -a(z), \quad v = -fx$$
$$p = -\frac{1}{2}f^2 x^2, \quad \rho = 0$$

The non-EB NCSF flow is horizontally unstable because its *G*-function depends on *z* only:

$$G(z) = -\frac{a'}{2a}$$

**(S12) Baroclinic elliptical or hyperbolic flow (Type VII)**

$$\psi = a(z)x^2 + b(z)y^2$$
$$u = -2by, \quad v = 2ax$$
$$p = 2ab(x^2+y^2) + f\psi$$
$$\rho = -\frac{f(a-b)'}{a-b}\psi, \quad \zeta = 2(a+b)$$

where for elliptical flow,

$$b = \frac{c - f \operatorname{arctanh}(h)}{2h}, \quad a = bh^2$$

and for hyperbolical flow,

$$b = \frac{c - f \arctan(h)}{2h}, \quad a = -bh^2$$

The two flows have the same $G$-function

$$G(z) = \frac{f(ab' - a'b)}{4ab(a-b)} = \frac{f h'}{2h(b-a)},$$

which depends on $z$ only. Therefore the flows are horizontally unstable.

**(S13)** Baroclinic parabolic flow (Type VII)

$$\psi = ax^2 + \frac{cay}{(f+2a)^2}$$

$$u = -\frac{ca}{(f+2a)^2}, \quad v = 2ax$$

$$p = fax^2 + \frac{cay}{f+2a}, \quad \rho = -\frac{fa'}{a}\psi$$

where $c$ is constant and $a = a(z)$. This non-EB baroclinic flow is horizontally unstable because its $G$-function depends on $z$ only:

$$G(z) = -\frac{fa'}{(f+2a)a}$$

The complete set of quadratic exact solutions, as classified by Table 2, displays rich topology and contains all stability types of Table 1. This stands in contrast to the limited geometry of high-degree polynomial solutions obtained by *Sun* (2016).

Table 2. Classification of quadratic flows (*italic* for baroclinic flows), with geometric stability corresponding to Table 1.

|  |  | Non-EB | EB |
|---|---|---|---|
| Non-rotating fluid | CSF | Straightline jet **(S1)** | Straightline jet **(S1)** <br> *Circular vortex* **(S2)** |
|  | NCSF | Elliptical vortex **(S3, S6)** <br> Hyperbolic flow **(S3, S5)** <br> Parabolic flow **(S4)** |  |
| Rotating Fluid | CSF | Inertial circular vortex with skew center **(S7)** | *Straightline jet* **(S8)** <br> *Circular vortex* **(S9)** |
|  | NCSF $G(x,y,z)$ | Inertial hyperbolic or elliptic flow **(S10)** |  |
|  | NCSF $G(z)$ | Inertial parabolic **(S11)** <br> *Elliptic vortex* **(S12)** <br> *Hyperbolic flow* **(S12)** <br> *Parabolic flow* **(S13)** |  |

## 5. Discussion

Geometric stability is an inherent property of ideal flows, because it is based on the exact solutions to the steady PIF model which does not involve boundary conditions. This differs from hydrodynamic stability which hinges on the temporal development of an externally perturbation. A laminar flow is said to be dynamically unstable with respect to certain perturbation, because different types of perturbation may excite different dynamic instabilities.

The two kinds of instability are intertwined in transient flows. A geometrically unstable flow is certain to incur dynamical instability, but a geometrically stable flow may not be dynamically stable. For example, a circular vortex with steep velocity profile is geometrically stable but could incur shear instability (*Flor and van Heijst* 1996). Another example comes from the classical baroclinic instability theory (*Pedlosky* 1987). The basic state of the f-plane Eady model is an inviscid baroclinic zonal flow similar to the pseudo-plane solution (S9). While this basic flow is geometrically stable, interaction between upper and lower boundaries induces unstable baroclinic wave with tilting vertical phase line (*Sun* 2007). Tilting phase line represents a distortion and loss of vertical coherence of the original laminar flow. It shows that a geometrically stable flow can be dynamically unstable under the effect of boundary.

An underlying assumption of geometric stability is that viscous attribution from boundary layers will cause non-proportional decay in the flow velocity field. The theory of geometric stability suggests that only EB flows, specifically straightline jet and circular vortex, can withstand viscous attribution and retain original streamline geometry. The experimental evidence is abundant: numerical simulation of elliptical vortex (*Dritschel and Juárez* 1996) and laboratory experiment of multipolar vortex (*Flor and van Heijst* 1996, *Billant and Chomaz* 2000) all show that vortices with non-circular cross-section are unstable and give rise to various instabilities. Experiment by *Boulanger et al.* (2008) further shows that a tilting circular vortex is unstable and instability occurs for tilt angles as small as 2° at high Reynolds numbers.

In laboratory and numerical experiments, vortices are known to have an intriguing tendency of axisymmetrization and vertical alignment (see references in Section I). The phenomenon can be explained by the geometric stability theorem, which states that the only geometrically stable flow with closed streamlines is vertically aligned circular vortex. Such low-dimensional flow structure represents a ground-state degeneracy that has been shown in topological studies to be robust against various perturbations.

As for the Boussinesq approximation applied in the PIF model, scale analysis shows that it is relatively accurate in flows with short scale height. In deep baroclinic flows the solenoidal term may not be small, but the geometric theory is still applicable if we simply define pseudo-plane flow as isobaric flow and adopt the pressure-coordinate laminar model in the form of Eqs.(A1-A5) of Sun08. There is no vorticity generation by pressure-density solenoidal in the isobaric system.

## 6. Conclusion

The geometric properties derived from a steady-state nonviscous Boussinesq-fluid model (named the PIF model) by *Sun* (2008) are extended to general pseudo-plane flows, revealing a fundamental relation between vertical coherence and streamline topology in ideal fluid. It suggests that for steady pseudo-plane flows, vertical alignment in terms of EB structure only appears in straightline jet and circular vortex.

In contrast with traditional hydrodynamic stability, the study introduces a concept of geometric stability for pseudo-plane flows based on the Gilbarg's problem: if an exact solution to the PIF model is non-unique, its geometric structure is stable and has high survivability in decaying turbulence; if an exact solution is unique, it is geometrically unstable and vulnerable to external perturbations.

The study shows all EB flows are geometrically stable and all non-straightline non-EB flows are unstable. Particularly, stable pseudo-plane flows with closed streamlines must assume vertical-aligned circular-vortex formulation, proffering a theoretical explanation for the phenomenon of vortex alignment and axisymmetrization.

**Acknowledgements.** This study has been supported by the National Natural Science Foundation of China (Grant No. 41576017)